\documentclass[twocolumn]{revtex4}[12pt]
\usepackage{graphicx,multirow,array,amsmath,empheq,amsthm, amssymb,amsfonts,tikz,ragged2e,color}
\begin{document}
\title{Self-gravitating envelope solitons in astrophysical compact objects}
\author{S. Khondaker$^*$, N. A. Chowdhury, A. Mannan, and A. A. Mamun}
\address{Department of Physics, Jahangirnagar University, Savar, Dhaka-1342, Bangladesh.\\
$^*$Email: khondakershohana03@gmail.com}
\begin{abstract}
  The propagation of ion-acoustic waves (IAWs) in a collisionless unmagnetized self-gravitating degenerate
  quantum plasma system (SG-DQPS) has been studied theoretically for the first time. A nonlinear Schr\"{o}dinger
  equation is derived by using the reductive perturbation method to study the nonlinear dynamics of the IAWs in the SG-DQPS.
  It is found that for $k_c > k$ ($k_c < k$) (where $k_c$ is critical value of the propagation constant $k$ which determines the stable and unstable region of IAWs)
  the IAWs are modulationally unstable (stable), and that $k_c$ depends only on the ratio of the electron number density  to light ion number density.
  It is also observed that the self-gravitating bright envelope solitons are modulationally stable. The results obtained from our present investigation
  are useful for understanding  the nonlinear propagation of the IAWs in astrophysical compact objects like white dwarfs and neutron stars.
\end{abstract}
\maketitle
\section{Introduction}
\label{1sec:int}
Recently, the self-gravity of degenerate quantum plasma (DQP) is the cornerstone  among the plasma
physicists to understand the basic features of the astrophysical compact objects (viz. white dwarf,
neutron stars \cite{Chandrasekhar1931,Fowler1994,Shapiro1983,Koester1990,Koester2002,Shukla2011,Zaman2017}) as
well as in laboratory environments (viz. solid density plasmas \cite{Drake2009,Drake2010}, laser
produced plasmas formed from sold targets irradiating by intense laser \cite{Glenzer2009}, ultra-cold
plasmas \cite{Fletcher2006, Killian2006}, etc.). The self-gravitating DQP system (SG-DQPS) has
a large number of  ultra-relativistic or non-relativistic degenerate species (order of $10^{30} cm^{-3}$
in white dwarfs, and  order of $10^{36} cm^{-3 }$ even more in neutron stars \cite{Shapiro1983,Koester1990,Koester2002}) and extremely
low temperature which exhibits unique collective behaviours from others plasma system.
The basic constituents of the SG-DQPS (viz. white dwarf, neutron stars) are
degenerate inertialess electron species \cite{Chandrasekhar1931, Fowler1994,Shapiro1983, Koester1990, Koester2002},
degenerate inertial light ion species (viz.${\rm ~^{1}_{1}H}$ \cite{, Fletcher2006, Killian2006}, ${\rm ~^{4}_{2}He}$
\cite{Chandrasekhar1931, Fowler1994}, and ${\rm ~^{12}_{~6}C}$ \cite{Koester1990, Koester2002}),
and heavy ion species (viz.${\rm ~^{56}_{ 26}Fe}$ \cite{Vanderburg2015}, ${\rm ~^{85}_{ 37}Rb}$ \cite{Witze2014}, and ${\rm ~^{96}_{ 42}Mo}$ \cite{Witze2014}).

The dynamics of the SG-DQPS is governed by the quantum mechanics because of the de Broglie
wavelength of particles is comparable to the inter-particle distance in SG-DQPS \cite{Shapiro1983, Koester1990}. According
to the Heisenberg's uncertainty principle, in quantum realm, the exact position and momentum of a particle
cannot be determined simultaneously, and mathematically it can be expressed
as $\Delta x\Delta p\ge \hbar/2$ (where $\Delta x$ is the uncertainty in position of the
particle and $\Delta p$ is the uncertainty in momentum of the same particle, and $\hbar$ is
the reduced Planck constant). In SG-DQPS, the position (momentum) of the plasma species
is well (not well) defined and these confined plasma species with uncertain momentum exerts
a pressure on the surrounding medium. Chandrasekhar more than 80 years ago defined this
exert pressure as degenerate pressure and mathematically it can be expressed as \cite{Chandrasekhar1931,Fowler1994}
\begin{eqnarray}
&&\hspace*{-4cm}P_j = K_j N_j^{\gamma},~~~~~K_j\simeq\frac{3}{5}\frac{\pi\hbar^2}{m_j},
\label{1eq:1}
\end{eqnarray}
where $j=e$ for the electron species, $j = l$ ($h$) for light (heavy) ion species,
$K_j$ is the proportional constant, $\gamma$ is a relativistic factor and $\gamma=5/4$
($5/3$) stands for ultra-relativistic (non-relativistic) limit, and $m_j$ is the mass
of the plasma species. The degenerate pressure of the SG-DQPS is dependent (independent) on the
number density and mass (temperature) of the plasma species. The mass of the plasma species generates
a strong gravitational field which provides the inward pull to compress the plasma system, but this inward pull
is counter-balanced by the outward degenerate pressure.

The amplitude of the ion-acoustic waves (IAWs) is appeared to modulation due to wave-particle
interaction,  the nonlinear self-interaction of the carrier wave modes, interaction between low
and high frequency modes \cite{Chowdhury2017a,Sultana2011}). The modulational instability (MI) and generation of the envelope
solitons in any nonlinear and dispersive medium are governed by the the nonlinear Schr\"{o}dinger (NLS)
equation. Recently, a large number of authors have studied the nonlinear wave propagation in
SG-DQPS. Asaduzzaman \textit{et al.} \cite{Asaduzzaman2017} have investigated the  nonlinear
propagation of self-gravitational perturbation mode in a super dense DQP medium. Mamun \cite{Mamun2017} analyzed
shock structures in a self-gravitating, multi-component DQP and found that the height and thickness of the
shock structures are totally dependent on the  dissipative and nonlinear coefficients.
Chowdhury \textit{et al.} \cite{Chowdhury2018} have reported that the MI of nucleus-acoustic waves (NAWs)
in a DQP system and found that the MI growth rate of the unstable NAWs is significantly modified by
the number density of nucleus species. Islam \textit{et al.} \cite{Islam2017} have studied envelope
solitons in three component DQP. However to the best of our knowledge, no attempt has been made
to study MI of the IAWs in SG-DQPS. Therefore, in the present work, we
will derive  a NLS equation by employing reductive perturbation method
to study the MI and formation of the envelope solitons in a SG-DQPS (containing inertialess
degenerate electron species, inertial degenerate light as well as heavy ion species).

The manuscript is organized as follows: The basic governing equations of our plasma
model is presented in Sec. \ref{1sec:gov}. Derivation of a NLS equation using reductive
perturbation technique is presented in Sec. \ref{1sec:der}. The stability of the IAWs and envelope solitons are examined
in Sec. \ref{1sec:MI }. A brief discussion is  provided in Sec. \ref{1sec:dis}.
\section{Governing Equations}
\label{1sec:gov}
We consider a SG-DQPS comprising of inertialess degenerate electron species $e$, inertial
degenerate light ion species $l$, and heavy ion species $h$, respectively. The detail
information about light and heavy nuclei is presented in Table \ref{Table:1}. The nonlinear
dynamics of such a SG-DQPS is governed by the following equations
\begin{eqnarray}
&&\hspace*{-2.7cm}\frac{\partial P_e}{\partial X}=-{m_e}{N_e}\frac{\partial \tilde{\psi}}{\partial X},
\label{1eq:2}\\
&&\hspace*{-2.7cm}\frac{\partial N_l}{\partial T}+\frac{\partial}{\partial X}(N_l U_l)=0,
\label{1eq:3}\\
&&\hspace*{-2.7cm}\frac{\partial U_l}{\partial T} + U_l\frac{\partial U_l }{\partial X}=-\frac{\partial \tilde{\psi}}{\partial X}-\frac{1}{m_lN_l} \frac{\partial P_l}{\partial X},
\label{1eq:4}\\
&&\hspace*{-2.7cm}\frac{\partial^2 \tilde{\psi}}{\partial X^2}= 4 \pi G~ [m_eN_e+m_lN_l+m_hN_h],
\label{1eq:5}
\end{eqnarray}
where $T(X)$ is the time (space) variable; $P_e$ ($P_l$) is the degenerate pressure associated
with degenerate electrons (light ions); $m_e$, $m_l$, and $m_h$ is the mass of electrons, light,
and heavy ions, respectively; $N_{e}$, $N_{l}$, and $N_{h}$ is, respectively, the number
densities of the electrons, light, and heavy ions; $U_l$ is the light ion fluid speed;
$\tilde{\psi}$ is the self-gravitational potential; $G$ is the universal gravitational constant.
Now, the quasi-neutrality condition at equilibrium can be expressed as
\begin{eqnarray}
&&\hspace*{-5.5cm}N_e = Z_l N_l + Z_h N_h,
\label{1eq:6}
\end{eqnarray}
where $Z_l$ ($Z_h$) is the charge state of a light (heavy) ion. For the purposes of simplicity,
we have considered the continuity and momentum balance equation for the inertial light ion species
$l$. Now, introducing normalized variables, specifically,
$x=X/L_q$, $t= T/\omega_{jl}$, $n_l = N_l/n_{l0}$, $u_l = U_l/C_q$, $\tilde{\psi} = C_q^{2}\psi$,
[where  $C_q = \sqrt{\pi}\hbar n_{e0}^{1/3}/m_l$, $\omega_{jl} = 4 \pi G m_l n_{lo}$, $L_{q} = C_q/\omega_{jl}$;
$n_{l0}$ ($n_{e0}$) is the equilibrium number densities of light ion species (electrons); $\psi$
is the dimensionless self-gravitational potential. After normalization, Eqs. (\ref{1eq:2})$-$(\ref{1eq:5})
appear in the following form
\begin{eqnarray}
&&\hspace*{-4cm}\frac{\partial \psi}{\partial x}=-\frac{3}{2}\alpha \frac{\partial n_e^{2/3}}{\partial x},
\label{1eq:7}\\
&&\hspace*{-4cm}\frac{\partial n_l}{\partial t}+\frac{\partial}{\partial x}(n_l u_l)=0,
\label{1eq:8}\\
&&\hspace*{-4cm}\frac{\partial u_l}{\partial t}+ u_l\frac{\partial u_l }{\partial x}=-\frac{\partial \psi}{\partial x}-\beta \frac{\partial n_l^{2/3}}{\partial x},
\label{1eq:9}\\
&&\hspace*{-4cm}\frac{\partial^2 \psi}{\partial x^2}= \gamma_e (n_e-1)-\gamma_l (n_l-1) ,
\label{1eq:10}
\end{eqnarray}
where $\alpha = m_l/m_e$, $\beta =(3/2)\mu^{-2/3}$, $\mu = n_{e0}/n_{l0}$, $\gamma_e = \mu(1/\alpha+\gamma/Z_l)$,
$\gamma_l = \gamma-1$; $\gamma = Z_lm_h/Z_hm_l$ (which is larger than 1 for any set of light
and heavy ion species). In $\gamma_e$, $1/\alpha \ll \gamma/Z_l$ (where $1/\alpha$
varies from $\sim 10^{-4}$ to $\sim 10^{-3}$, and $\gamma/Z_l$ varies from $\sim 0.1$ to $\sim 2.0$),
so $1/\alpha$ is negligible compared to $\gamma/Z_l$, and can be written as
$\gamma_e \simeq \mu \gamma/Z_l$. For inertialess degenerate electron species, the expression for the number density is
\begin{eqnarray}
&&\hspace*{-2cm}n_e=\left[1-\frac{2\psi}{3 \alpha^2}\right]^{\frac{3}{2}}
\nonumber\\
&&\hspace*{-1.5cm}=1-\frac{1}{\alpha^2} \psi+ \frac{1}{6\alpha^4}\psi^2 +\frac{1}{54\alpha^6} \psi^3+ \cdot \cdot \cdot \cdot.
\label{1eq:11}
\end{eqnarray}
Now, by substituting Eq. (\ref{1eq:11}) into Eq. (\ref{1eq:10}), and expanding up to third order in $\psi$, we get
\begin{eqnarray}
&&\hspace*{-1.5cm}\frac{\partial^2 \psi}{\partial x^2}-\gamma_l+\gamma_l n_l=\gamma_1\psi+\gamma_2 \psi^2+\gamma_3 \psi^3+\cdot\cdot\cdot\cdot,
\label{1eq:12}
\end{eqnarray}
where
$\gamma_1= - \gamma_e/\alpha^2$,  $\gamma_2=\gamma_e/6\alpha^4$, and  $\gamma_3=\gamma_e/54\alpha^6$.
We note that the terms on the right hand side of Eq. (\ref{1eq:12}) is the contribution of electron.
\begin{center}
\begin{table}[h!]
\centering
\caption{The values of $\gamma$ when ${\rm ~^{1}_{1}H}$ \cite{Fletcher2006,Killian2006},
${\rm ~^{4}_{2}He}$ \cite{Chandrasekhar1931,Fowler1994}, and  ${\rm ~^{12}_{~6}C}$
\cite{Koester1990,Koester2002} are considered as the light ion species,  and ${\rm ~^{56}_{ 26}Fe}$
\cite{Vanderburg2015},  ${\rm ~^{85}_{ 37}Rb}$ \cite{Witze2014}, and  ${\rm ~^{96}_{42}Mo}$ \cite{Witze2014}
are considered as the heavy ion species.}
\begin{tabular}{|p{3cm}|p{2.5cm}|m{1.5cm}|}
\hline
{~~~~~\bf Light ion}                                         &~~~~{\bf Heavy ion}                                                             &~~~~~${\bf \gamma}$ \\ [2.2ex]
\hline
 \multirow{2}{4em}{~}                                   & ~~~~${\rm ~^{56}_{ 26}Fe}$  \cite{Vanderburg2015}                                   &~~~~2.16 \\ [2.2ex]
\cline{2-3}

${\rm ~~~~~~^{1}_{1}H}$ \cite{Fletcher2006,Killian2006}      &~~~~${\rm ~^{85}_{ 37}Rb}$  \cite{Witze2014}                                          &~~~~2.30\\[2.2ex]
\cline{2-3}
 ~                                                      & ~~~~${\rm ~^{96}_{42}Mo}$  \cite{Witze2014}                                              &~~~~2.28\\[2.2ex]
\hline
 \multirow{2}{5em}{~}                            & ~~~~${\rm ~^{56}_{ 26}Fe}$  \cite{Vanderburg2015}                                             &~~~~1.08 \\[2.2ex]
\cline{2-3}

${\rm ~~~~~~^{4}_{2}He}$ \cite{Chandrasekhar1931,Fowler1994}   &~~~~${\rm ~^{85}_{ 37}Rb}$  \cite{Witze2014}                                    &~~~~1.15\\[2.2ex]
\cline{2-3}
 ~                                                                    & ~~~~${\rm ~^{96}_{42}Mo}$  \cite{Witze2014}                          &~~~~1.14\\[2.2ex]
\hline
 \multirow{2}{*}{~}                                    & ~~~~${\rm ~^{56}_{ 26}Fe}$ \cite{Vanderburg2015}                                       &~~~~1.08 \\[2.2ex]
\cline{2-3}

${\rm ~~~~~~^{12}_{~6}C}$ \cite{Koester1990,Koester2002}   &~~~~${\rm ~^{85}_{ 37}Rb}$ \cite{Witze2014}                                        &~~~~1.15\\[2.2ex]
\cline{2-3}
 ~                                                 & ~~~~${\rm ~^{96}_{42}Mo}$   \cite{Witze2014}                                             &~~~~1.14\\[2.2ex]
\hline
\end{tabular}
\label{Table:1}
\end{table}
\end{center}
\section{Derivation of the NLS Equation}
\label{1sec:der}
\begin{figure}[t!]
  \centering
  \includegraphics[width=80mm]{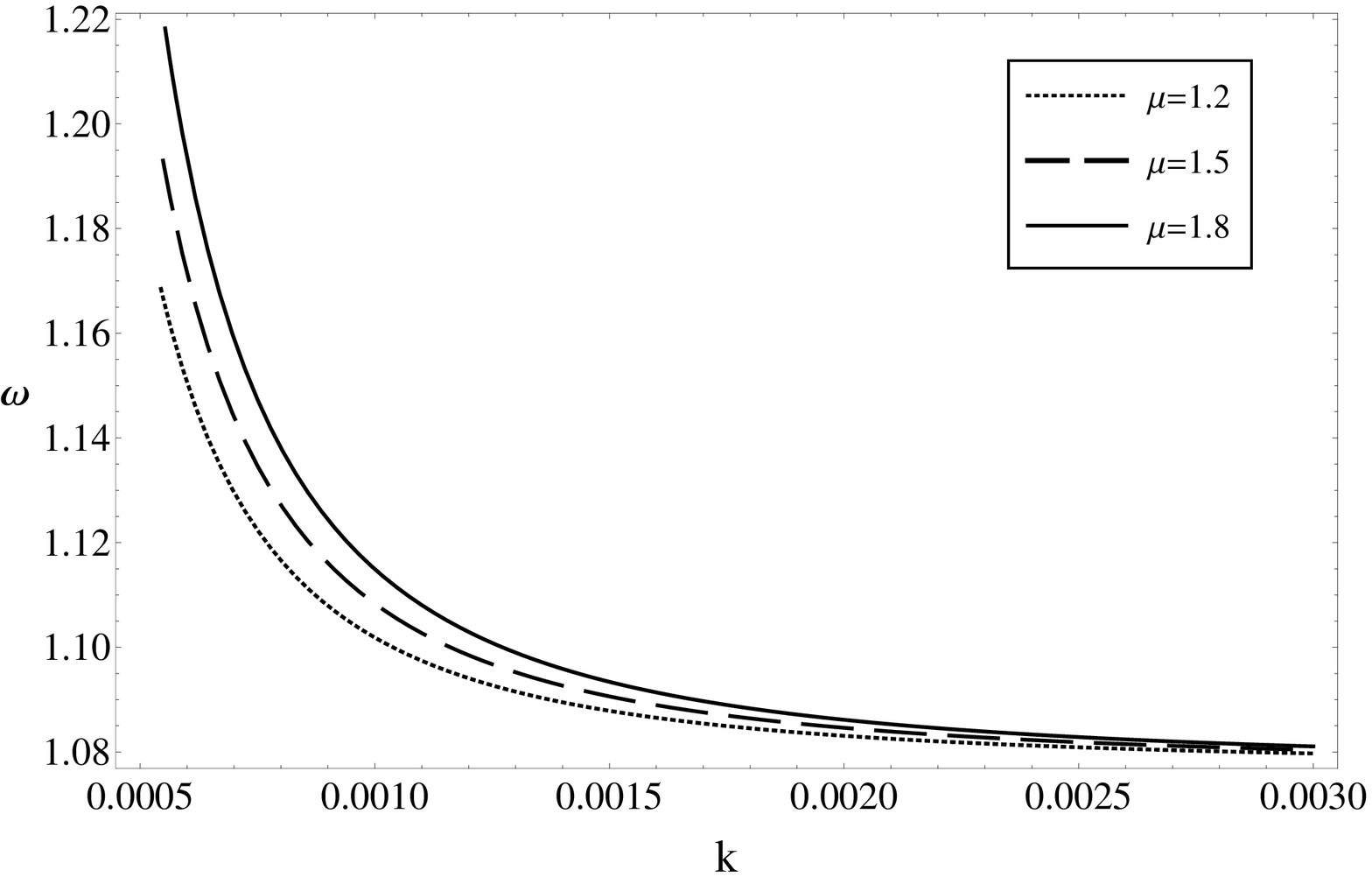}
  \caption{The variation of $\omega$ with $k$ for different values
  of $\mu$; along with $\alpha=3.67\times10^3$, $\gamma=2.16$, and $\gamma/Z_l=0.5$.}
  \label{1Fig:F1}
  \vspace{0.8cm}
  \includegraphics[width=90mm]{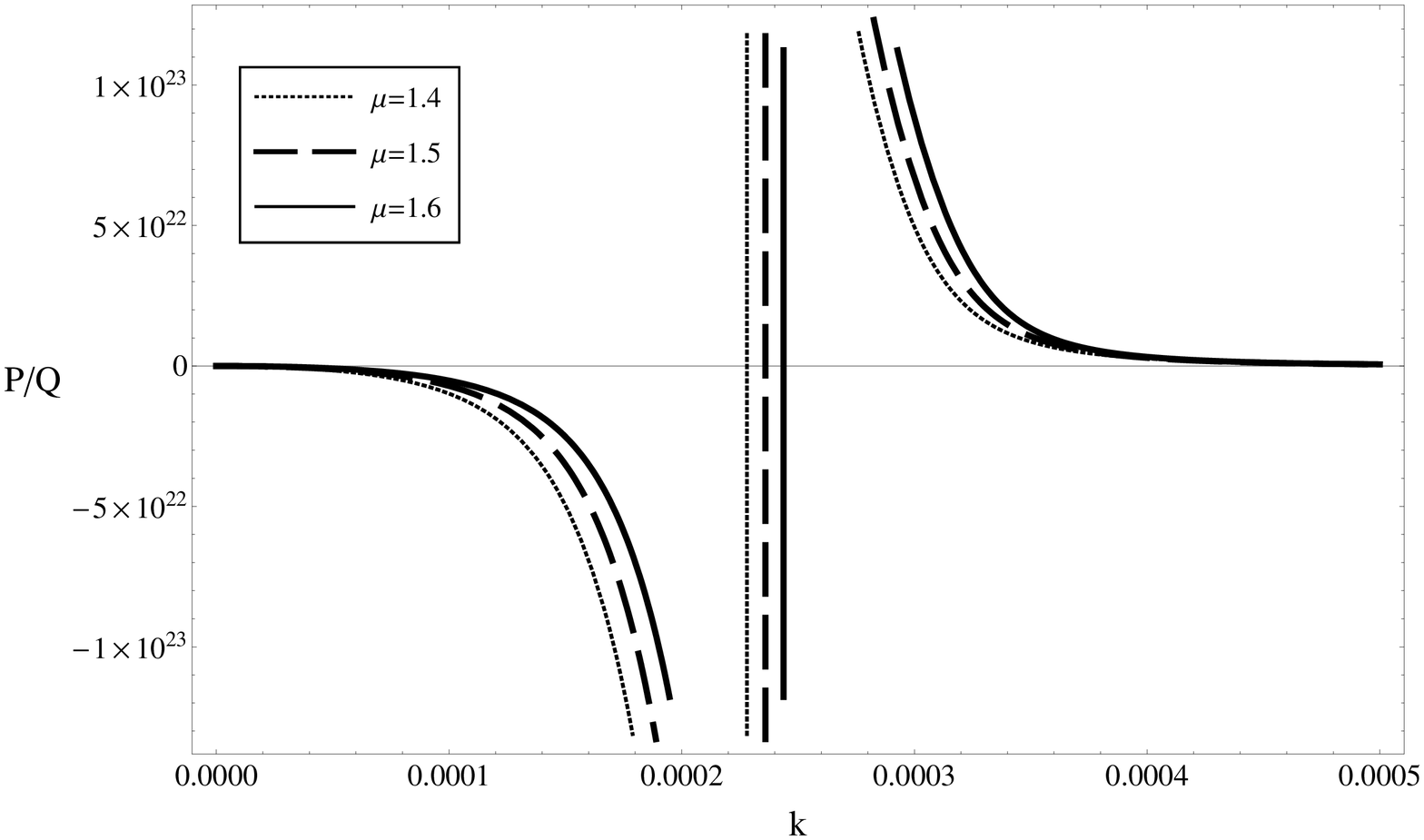}
  \caption{The variation of $P/Q$ with $k$ for different values of
  $\mu$; along with $\alpha=3.67\times10^3$, $\gamma=2.16$, and $\gamma/Z_l=0.5$.}
  \label{1Fig:F2}
\end{figure}

In order to demonstrate the MI and the basic features of IAWs in a SG-DQPS, we employ the standard
reductive perturbation \cite{Taniuti1969,Chowdhury2017b} method in which independent variables are stretched as
\begin{equation}
\left.
\begin{array}{l}
\hspace*{-3.7cm}\xi=\epsilon(x-v_gt),\\
\hspace*{-3.7cm}\tau=\epsilon^2t,
\end{array}
\right\}
\label{1eq:13}
\end{equation}
hence, we have
\begin{eqnarray}
&&\hspace*{-3.8cm}\frac{\partial}{\partial t}\rightarrow\frac{\partial}{\partial t}-\epsilon v_g \frac{\partial}{\partial\xi}+\epsilon^2\frac{\partial}{\partial\tau},
\label{1eq:14}\\
&&\hspace*{-3.8cm}\frac{\partial}{\partial x}\rightarrow\frac{\partial}{\partial
x}+\epsilon\frac{\partial}{\partial\xi},
\label{1eq:15}
\end{eqnarray}
where $\epsilon$ is a small parameter and $v_g$ is the real variable interpreted as the group velocity.
Furthermore, the dependent variables $n_l$, $u_l$, and  $\psi$ can be expanded in power series of $\epsilon$ as
\begin{eqnarray}
&&\hspace*{-1.3cm}n_{l}=1+\sum_{m=1}^{\infty}\epsilon^{(m)}\sum_{l'=-\infty}^{\infty}n_{ll'}^{(m)}(\xi,\tau)~\mbox{exp}[il'\Upsilon],
\label{1eq:16}\\
&&\hspace*{-1.3cm}u_{l}=\sum_{m=1}^{\infty}\epsilon^{(m)}\sum_{l'=-\infty}^{\infty}u_{ll'}^{(m)}(\xi,\tau)~\mbox{exp}[il'\Upsilon],
\label{1eq:17}\\
&&\hspace*{-1.3cm}\psi=\sum_{m=1}^{\infty}\epsilon^{(m)}\sum_{l'=-\infty}^{\infty}\psi_{l'}^{(m)}(\xi,\tau)~\mbox{exp}[il'\Upsilon],
\label{1eq:18}
\end{eqnarray}
where $\Upsilon=(kx-\omega t)$ and $\omega$ ($k$)  corresponds to the angular frequency
(wave number) of the carrier waves, respectively. Now, by replacing the Eqs.
(\ref{1eq:13})$-$(\ref{1eq:18}) into Eqs. (\ref{1eq:8}), (\ref{1eq:9}), and (\ref{1eq:12}),
and collecting all terms of similar power of $\epsilon$, the first order ($m=1$ with
$l'=1$) reduced equations can be represented as
\begin{eqnarray}
&&\hspace*{-6cm}n_{l1}^{(1)}=\frac{k^2}{S}\psi_1^{(1)},
\label{1eq:19}\\
&&\hspace*{-6cm}u_{l1}^{(1)}=\frac{k \omega}{S}\psi_1^{(1)},
\label{1eq:20}
\end{eqnarray}
where $S=\omega^2-\beta_1 k^2$ and $\beta_1=2\beta/3$. The linear dispersion relation can be obtained
from the first-order equations in the form
\begin{eqnarray}
&&\hspace*{-5cm}\omega^2=\frac{\gamma_l k^2}{k^2+\gamma_1}+\beta_1 k^2.
\label{1eq:21}
\end{eqnarray}
The dispersion properties of IAWs for different values of $\mu$ is depicted
in Fig. \ref{1Fig:F1} and it may deduce that (a) the value of $\omega$  exponentially decreases
with the increase of $k$; (b) on the other hand, the value of $\omega$
increases (decreases) with $n_{e0}$ ($n_{l0}$).
Next, the second-order ($m=2$ with $l'=1$) reduced equations are given by
\begin{eqnarray}
&&\hspace*{-1.5cm}n_{l1}^{(2)}=\frac{k^2}{S}\psi_1^{(2)}+\frac{2i\omega k(v_g k-\omega)}{S^2} \frac{\partial \psi_1^{(1)}}{\partial\xi},
\label{1eq:22}\\
&&\hspace*{-1.5cm}u_{l1}^{(2)}=\frac{k \omega}{ S}\psi_1^{(2)} +\frac{i (\beta_1 k^2+\omega^2 )(v_g k-\omega)}{S^2} \frac{\partial\psi_1^{(1)}}{\partial\xi},
\label{1eq:23}
\end{eqnarray}
with the compatibility condition
\begin{eqnarray}
&&\hspace*{-3.5cm}v_g=\frac{\partial \omega}{\partial k}=\frac{\gamma_l\omega^2-(\omega^2-\beta_1 k^2 )^2}{\gamma_l k\omega}.
\label{1eq:24}
\end{eqnarray}
The amplitude of the second-order harmonics is found to be proportional to $|\psi_1^{(1)}|^2$
\begin{equation}
\left.
\begin{array}{l}
\hspace*{-1.8cm}n_{l2}^{(2)}=C_1|\psi_1^{(1)}|^2,~~~~~~~~~n_{l0}^{(2)}=C_4|\psi_1^{(1)}|^2,\\
\hspace*{-1.8cm}u_{l2}^{(2)}=C_2 |\psi_1^{(1)}|^2,~~~~~~~~~u_{l0}^{(2)}=C_5|\psi_1^{(1)}|^2,\\
\hspace*{-1.8cm}\psi_2^{(2)}=C_3 |\psi_1^{(1)}|^2,~~~~~~~~~\psi_0^{(2)}=C_6|\psi_1^{(1)}|^2, \\
\end{array}
\right\}
\label{1eq:25}
\end{equation}
where
\begin{eqnarray}
&&\hspace*{-1.0cm}C_1=\frac{3 \omega^2 k^4+ 2 C_3 k^2 S^2 }{2 S^2},
\nonumber\\
&&\hspace*{-1cm}C_2=\frac{C_1 \omega S^2 - \omega k^4}{k S^2},
\nonumber\\
&&\hspace*{-1cm}C_3=\frac{3 \gamma_l \omega^2 k^4 - 2\gamma_2 S^3}{2 S^3(4k^2+\gamma_1)-2 \gamma_l k^2S^2},
\nonumber\\
&&\hspace*{-1cm}C_4=\frac{2 \omega v_g k^3- \beta_2 k^4 + k^2\omega^2 + C_6 S^2 }{S^2(v_g^2- \beta_1)},
\nonumber\\
&&\hspace*{-1cm}C_5=\frac{C_4 v_g S^2 - 2  \omega k^3 }{S^2 },~~\beta_2 =\frac{\beta}{9}
\nonumber\\
&&\hspace*{-1cm}C_6=\frac{\gamma_l(2 \omega v_g k^3- \beta_2 k^4 + k^2\omega^2) -2 \gamma_2 S^2(v_g^2- \beta_1)}{\gamma_1 S^2(v_g^2- \beta_1) - \gamma_l S^2  }.
\nonumber
\end{eqnarray}
Finally, the third harmonic modes ($m=3$ with $l'=1$) provide a set of equations and after some
mathematical calculation these equations reduce [with the help of  Eqs. (\ref{1eq:19})$-$(\ref{1eq:25})]
to the following  NLS equation:
\begin{eqnarray}
&&\hspace*{-4cm}i\frac{\partial \Phi}{\partial \tau}+P\frac{\partial^2\Phi}{\partial \xi^2}+Q|\Phi|^2\Phi=0,
\label{1eq:26}
\end{eqnarray}
where $\Phi=\psi_1^{(1)}$ for simplicity. The dispersion coefficient $P$ and the nonlinear coefficient $Q$ are given by
\begin{eqnarray}
&&\hspace*{-2cm}P=\frac{v_g \beta_1^2 k^5 + 4 \beta_1 k^2 \omega^3 + 2 \beta_1 v_g \omega^2 k^3 - F_1}{2 \gamma_l k^2 \omega^2},
\label{1eq:27}
\end{eqnarray}
\begin{eqnarray}
&&\hspace*{-1cm}Q=\frac{1}{2 \gamma_l \omega k^2 S^2 }\left[2 \gamma_2 (C_3+C_6)S^4 + 2 \gamma_3 S^4 -F_2 \right],
\label{1eq:28}
\end{eqnarray}
where $F_1 = 3 v_g k \omega^4 +4 \omega \beta_1^2 k^4$, $ F_2=\gamma_l k^2\omega^2 S^2 (C_1+C_4)+2
\gamma_l\omega S^2 k^3 (C_2+C_5)+\gamma_l \beta_3 k^8$, and $\beta_3=4\beta/81$.
\section{MI and envelope solitons }
\label{1sec:MI }
\begin{figure}[t!]
  \centering
  \includegraphics[width=80mm]{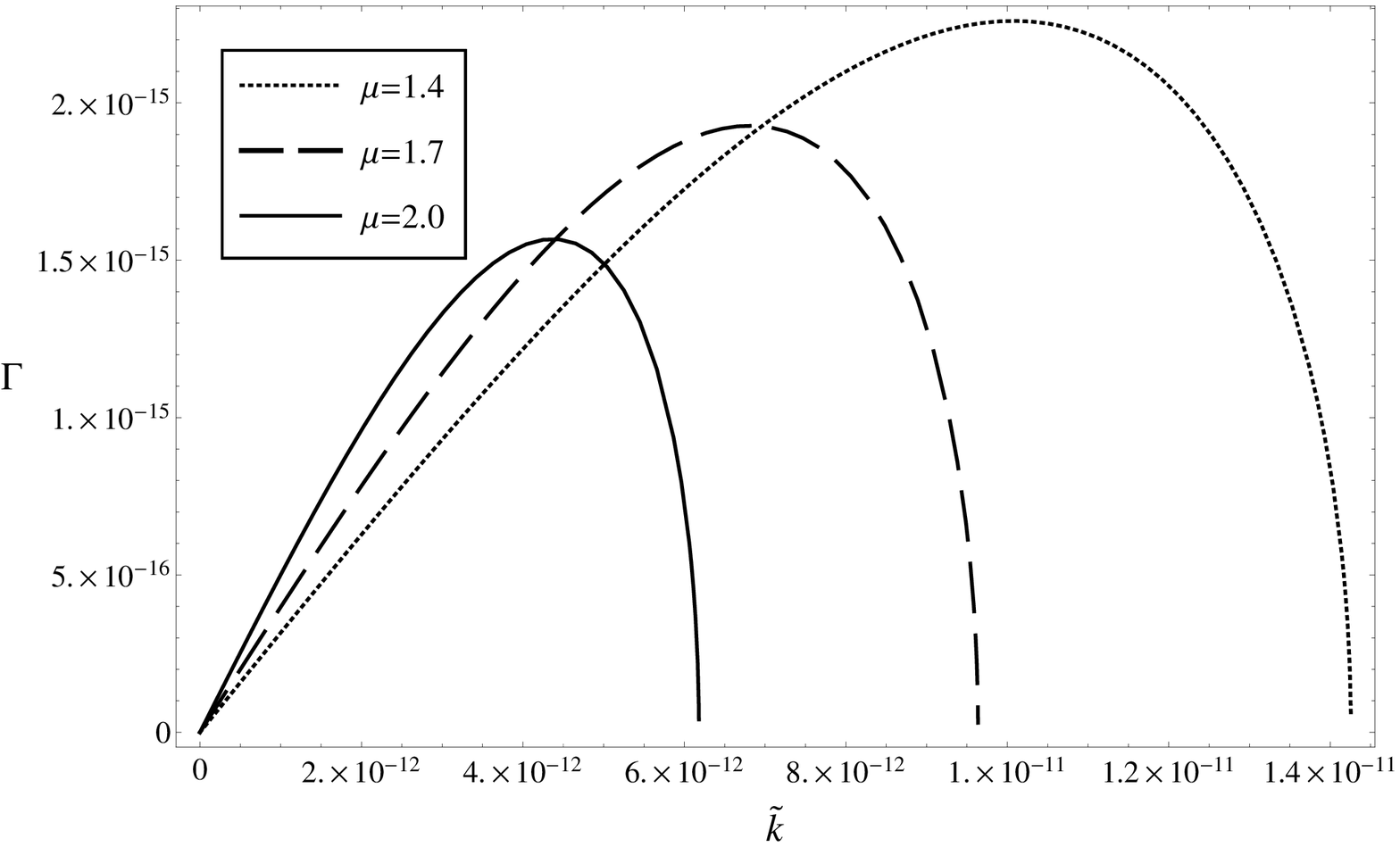}
  \caption{The variation of $\Gamma$ with ${\tilde{k}}$ for different values of $\mu$; along
  with $\alpha=3.67\times10^3$, $\gamma=2.16$, $\gamma/Z_l=0.5$, $k=0.00035$, and $\phi_0=0.9$.}
  \label{1Fig:F3}
  \vspace{0.8cm}
  \includegraphics[width=80mm]{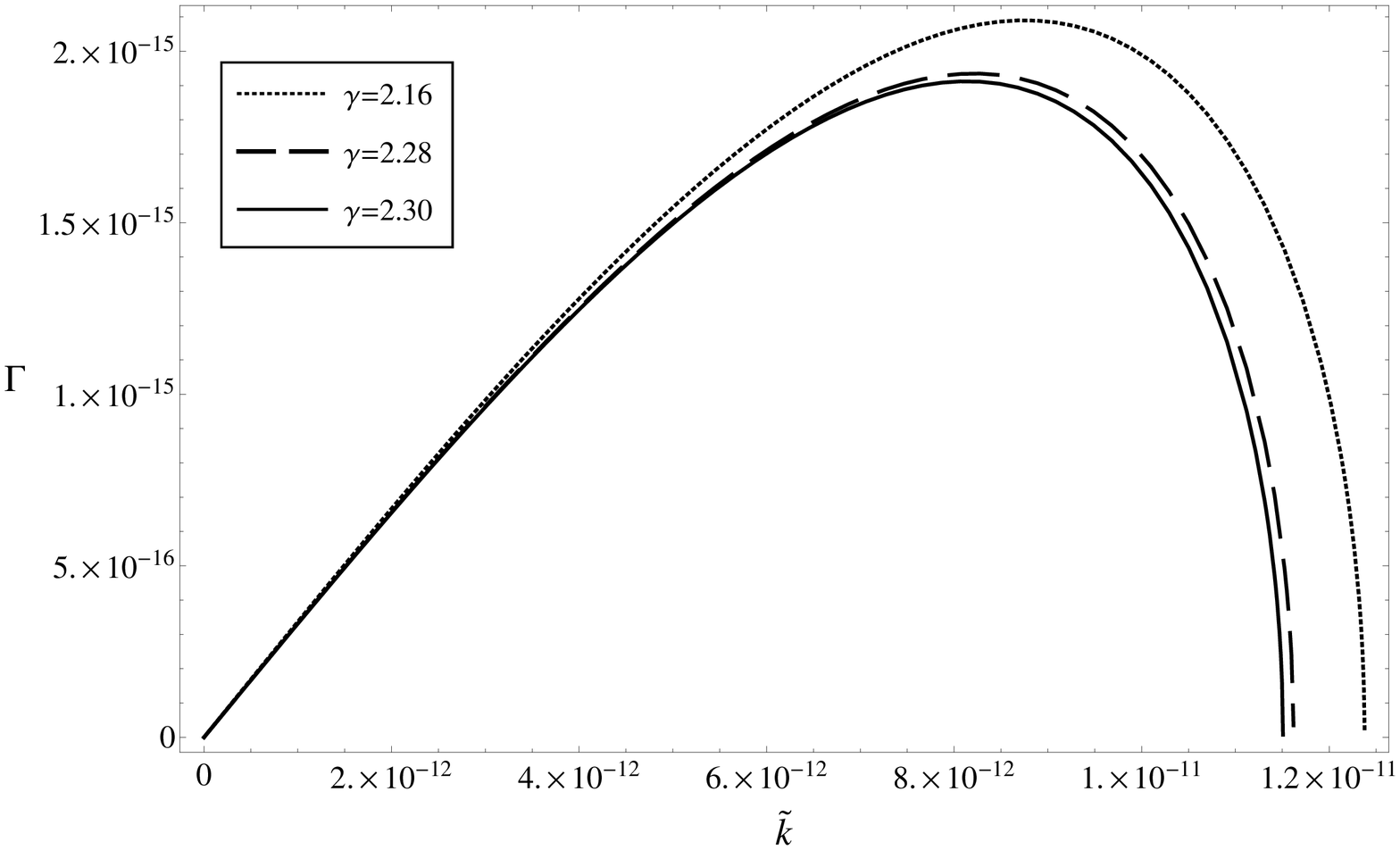}
  \caption{The variation of $\Gamma$
  with ${\tilde{k}}$ for different values of $\gamma$;
  along with $\alpha=3.67\times10^3$, $\gamma/Z_l=0.5$, $\mu=1.5$, $k=0.00035$, and $\phi_0=0.9$.}
  \label{1Fig:F4}
\end{figure}

The MI of IAWs can be studied by considering the harmonic modulated amplitude
solution of Eq. (\ref{1eq:26}) of the form  $\Phi=\hat{\Phi}e^{i\,Q|\hat{\Phi}|^2\tau} + c.~c.$ (c. c. being the complex conjugate),
where perturbed amplitudes are $\hat{\Phi} = \hat{\Phi}_0 + \epsilon \hat{\Phi}_1$ and
$\hat{\Phi}_1=\hat{\Phi}_{1,0}\exp[i(\tilde{k}\xi-\tilde{\omega}\tau)]+c.~c$
(here, the perturbed wave number $\tilde{k}$ and the frequency $\tilde{\omega}$ are different from $k$
and $\omega$). Hence, the nonlinear dispersion relation for the amplitude modulation obtained
by substituting these values in Eq. (\ref{1eq:26}) can be written
as \cite{Sultana2011,Schamel2002,Chowdhury2018,Kourakis2005,Fedele2002}
\begin{eqnarray}
&&\hspace*{-4.3cm}{\tilde{\omega}^2}=P^2{\tilde{k}^2} \left({\tilde{k}^2}-\frac{2{|\hat{\Phi}_o|^2}}{P/Q}\right).
\label{1eq:29}
\end{eqnarray}
It is apparent from Eq. (\ref{1eq:29}) that the IAWs will be modulationally stable (unstable) for the
range of values of $\tilde{k}$ in which $P/Q$ is negative (positive), such as,
$P/Q <0$ ($P/Q > 0$). When $P/Q\rightarrow\pm\infty$, the corresponding value of $k$
($=k_c$) is called threshold or critical wave number for the onset of MI. This $k_c$ separates the  unstable region
($P/Q>0$) from the stable ($P/Q<0$) one. The stability of the  profile has been investigated
by depicting the ratio of $P/Q$ with carrier wave number $k$ for different values of $\mu$  in  Fig. \ref{1Fig:F2},
which clearly indicate that (a) for the large (small) $k$, there is an unstable (stable) region
for IAWs; (b)  the $k_c$ increases (decreases) with the increase of the value of $n_{e0}$ ($n_{lo}$). So, the
electron and ion number densities play an opposite role to recognize the stability domain of the IAWs.
In the unstable region $P/Q>0$ and under this certain condition $\tilde{k}<k_c=\sqrt{2Q{|\hat{\Phi}_o|}^2/P}$,
the growth rate ($\Gamma$) of  MI is obtained from the Eq. (\ref{1eq:29}) can be written as
\begin{eqnarray}
&&\hspace*{-5.3cm}\Gamma=|P|~{\tilde{k}^2}\sqrt{\frac{k^2_{c}}{\tilde{k}^2}-1},
\label{1eq:30}
\end{eqnarray}
where $\hat{\Phi}_o$ is the amplitude of the carrier waves.
We have numerically analysed the influence of different plasma parameters on the MI growth rate by
depicting  $\Gamma$ with $\tilde{k}$ [obtained from Eq.(\ref{1eq:30})] for different values
of $\mu$ and $\gamma$ in Figs. \ref{1Fig:F3} and \ref{1Fig:F4} and  it is obvious  that (a) initially,
the $\Gamma$ increases with $\tilde{k}$ before obtained  it's maximum value $\Gamma_{max}$. But after
$\Gamma_{max}$, the $\Gamma$ decreases to zero for further increase in $\tilde{k}$; (b) as we increase the value of the
electron number density ($n_{e0}$), the maximum value of the growth rate decreases but increases with increases of the
light ion number density ($n_{l0}$); (c) on the other hand, the
growth rate $\Gamma$ decreases  with the increase (decrease) of $m_h$ ($m_l$) for
fixed value of $Z_l$ and $Z_h$ (via $\gamma=Z_lm_h/Z_hm_l$); (d) similarly, $\Gamma$ decreases  with the increase (decrease) of $Z_l$ ($Z_h$) for
fixed value of $m_h$ and $m_l$ (via $\gamma$). The physics of this result is that the maximum value of the growth rate increases as the nonlinearity of the plasma system
increases  with the increase (decrease) of the value of $Z_h$ or $m_l$ ($Z_l$ or $m_h$).

The self-gravitating bright solitons are generated when the carrier wave is modulationally unstable in the region $P/Q>0$, whose general analytical
form  reads as \cite{Sultana2011,Schamel2002,Chowdhury2018,Kourakis2005,Fedele2002}
\begin{eqnarray}
&&\hspace*{-2.3cm}\Phi(\xi,\tau)=\left[\psi_{0}~\mbox{sech}^{2}\left(\frac{\xi-U\tau}{W}\right)\right]^{1/2}\nonumber\\
&&\hspace*{-1.17cm}\times\exp\left[\frac{i}{2P} \left\{U\xi+\left(\Omega_{0}-\frac{U^2}{2}\right)\tau\right\}\right],
\label{1eq:31}
\end{eqnarray}
\begin{figure}[t!]
 \includegraphics[width=80mm]{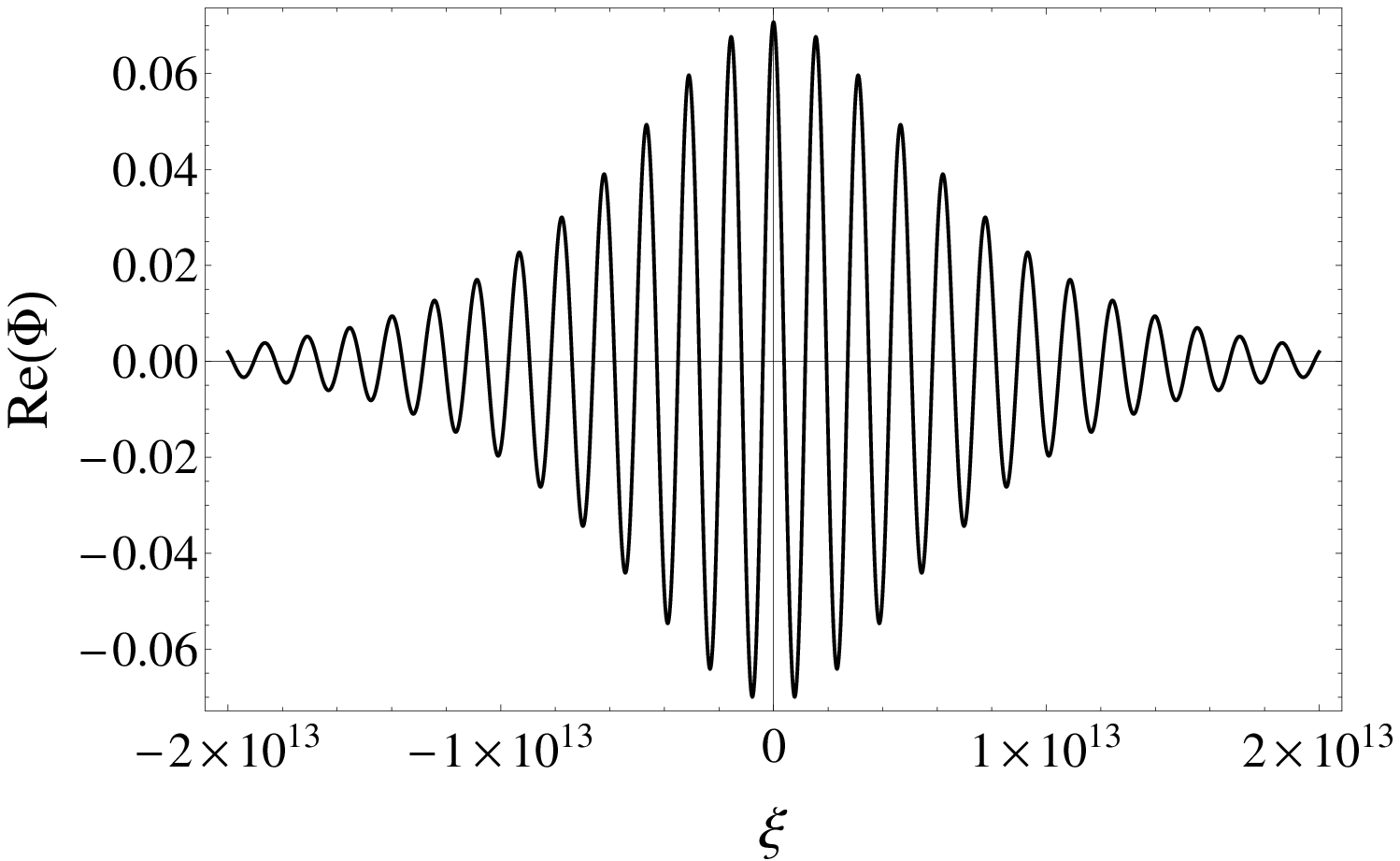}
  \caption{The variation of the $|\Phi|$ with $\xi$ for self-gravitating bright solitons; along with
  $\alpha=3.67\times10^3$, $\gamma= 2.16$, $\mu=1.5$, $k=0.0003$, $\psi_0=$0.005, $U=0.001$, $\Omega_0=0.04$, $\tau=0$. }
  \label{1Fig:F5}
  \vspace{0.8cm}
  \includegraphics[width=80mm]{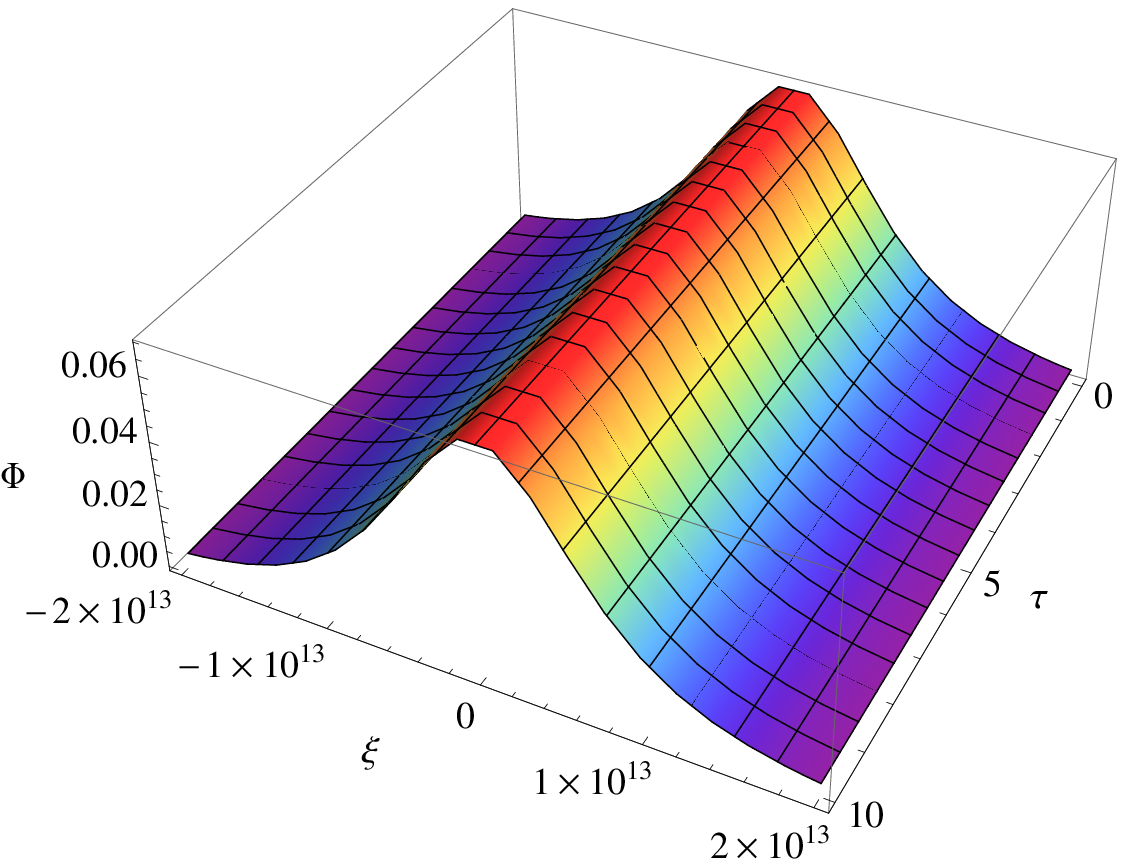}
  \caption{The variation of the $|\Phi|$ with $\xi$ and $\tau$ for self-gravitating bright solitons; along with
  $\alpha=3.67\times10^3$, $\gamma= 2.16$, $\mu=1.5$, $k=0.0003$, $\psi_0=$0.005, $U=0.001$, $\Omega_0=0.04$, $\tau=0$. }
  \label{1Fig:F6}
\end{figure}
where $U$ is the propagation speed, $\psi_0$ is the envelope amplitude, $\Omega_0$ oscillating frequency for $U=0$
and $W$ is the soliton width which can be defined as $W=\sqrt{2|P/Q|/ \psi_{0}}$. The self-gravitating bright envelope soliton
depicted in Figs \ref{1Fig:F5} and \ref{1Fig:F6}, clearly indicates that the shape of the self-gravitating bright envelope solitons is not affected
by any external perturbation through the time evolution. So, the self-gravitating bright envelope solitons are modulationally
stable.
\section{Discussion}
\label{1sec:dis}
In this work, we have investigated the amplitude modulation of IAWs in an unmagnetized three components SG-DQPS comprising of
inertialess degenerate electron species, inertial degenerate light  and heavy ion species.
A NLS equation is derived by employing the reductive perturbation method that governs the stability,
circumstance for the appearance of MI growth rate, and formation of the IAWs envelope solitons in SG-DQPS.
The noticeable results found from this theoretical investigation can be outlined as follows:
\begin{enumerate}
\item{The $\omega$ increases (decreases) with $n_{e0}$ ($n_{l0}$), and also decreases exponentially with the increase of $k$.}
\item{The IAWs will be stable (unstable) for  smaller values of $k$ and $P/Q < 0$ (larger values of $k$ and $P/Q > 0$).}
\item{The growth rate decreases with the increase of $n_{e0}$ but it decreases with increase (decrease) of $m_h$ ($m_l$) for fixed value of $Z_l$ and $Z_h$,
 similarly, growth rate decreases  with the increase (decrease) of $Z_l$ ($Z_h$) for fixed value of $m_h$ and $m_l$ (via $\gamma= Z_lm_h/Z_hm_l$).}
\item{The shape of the self-gravitating bright envelope solitons is not affected by any external perturbation through the
     time evolution. So, the self-gravitating bright envelope solitons are modulationally stable.}
\end{enumerate}
In conclusion, we hope that the results from our present theoretical investigation may be helpful in understanding the nonlinear
phenomena in astrophysical compact objects (viz. white dwarf and neutron stars \cite{Chandrasekhar1931,Fowler1994,Shapiro1983,Koester1990,Koester2002,Shukla2011}).
\section*{Acknowledgement}
S. Khondaker is thankful to the Bangladesh Ministry of Science and Technology for
awarding the National Science and Technology (NST) Fellowship.


\begin{thebibliography}{0}
\expandafter\ifx\csname natexlab\endcsname\relax\def\natexlab#1{#1}\fi
\expandafter\ifx\csname bibnamefont\endcsname\relax
  \def\bibnamefont#1{#1}\fi
\expandafter\ifx\csname bibfnamefont\endcsname\relax
  \def\bibfnamefont#1{#1}\fi
\expandafter\ifx\csname citenamefont\endcsname\relax
  \def\citenamefont#1{#1}\fi
\expandafter\ifx\csname url\endcsname\relax
  \def\url#1{\texttt{#1}}\fi
\expandafter\ifx\csname urlprefix\endcsname\relax\def\urlprefix{URL }\fi
\providecommand{\bibinfo}[2]{#2}
\providecommand{\eprint}[2][]{\url{#2}}

\end{thebibliography}


\begin{thebibliography}{99}

\bibitem{Chandrasekhar1931} S. Chandrasekhar, Astrophys. J. \textbf{74}, 81 (1931).

\bibitem{Fowler1994} R. H. Fowler, J. Astrophys. Astr. \textbf{15}, 115 (1994).

\bibitem{Shapiro1983} S. L. Shapiro and S. A. Teukolsky, \textit{Black Holes, White Dwarfs and Neutron Stars: The Physics of Compact Objects} (John Wiley \& Sons, New York, 1983).

\bibitem{Koester1990} D. Koester and G. Chanmugam, Rep. Prog. Phys. \textbf{53}, 837 (1990).

\bibitem{Zaman2017} D. M. S. Zaman, M. Amina, P. R. Dip, and A. A. Mamun, Eur. Phys. J. Plus \textbf{132}, 457 (2017).

\bibitem{Koester2002} D. Koester, Astron.  Astrophys. Rev. \textbf {11}, 33 (2002).

\bibitem{Shukla2011} P. K. Shukla and B. Eliasson, Rev. Mod. Phys. \textbf{83}, 885 (2011).

\bibitem{Drake2009} R. P. Drake, Phys. Plasmas \textbf{16}, 055501 (2009).

\bibitem{Drake2010} R. P. Drake, Phys. Plasmas \textbf{63}, 28 (2010).

\bibitem{Glenzer2009} S. H. Glenzer and  R. Redmer, Rev. Mod. Phys. \textbf{81}, 1625 (2009).

\bibitem{Fletcher2006} R. S. Fletcher, X. L. Zhang, and S. L. Rolston, Phys. Rev. Lett.  \textbf{96}, 105003 (2006).

\bibitem{Killian2006} T. C. Killian, Nature (London). \textbf {441}, 297 (2006).

\bibitem{Vanderburg2015} A. Vanderburg, J. A. Johnson, S. Rappaport, A. Bieryla, J. Irwin, J. A. Lewis, D. Kipping, W. R. Brown, P. Dufour, D. R. Ciardi, R. Angus,
                         L. Schaefer, D. W. Latham, D. Charbonneau, C. Beichman, J. Eastman, N. McCrady, R. A. Wittenmyer, and J. T. Wright, Nature \textbf{526}, 546 (2015).

\bibitem{Witze2014} A. Witze, Nature \textbf{510}, 196 (2014).

\bibitem{Chowdhury2017a} N. A. Chowdhury, A. Mannan, M. M. Hasan, and A. A. Mamun, Chaos \textbf{27}, 093105 (2017).

\bibitem{Sultana2011} S. Sultana and I. Kourakis, Plasma Phys. Control. Fusion \textbf{53}, 045003 (2011).

\bibitem{Asaduzzaman2017} M. Asaduzzaman, A. Mannan, A.A. Mamun, Phys. Plasmas \textbf{24}, 052102 (2017).

\bibitem{Mamun2017} A. A. Mamun, Phys. Plasmas \textbf{24}, 102306 (2017).

\bibitem{Chowdhury2018} N. A. Chowdhury, M. M. Hasan, A. Mannan, and A. A. Manun, Vacuum  \textbf{147}, 31 (2018).

\bibitem{Islam2017}  S. Islam, S. Sultana, and A. A. Mamun, Phys. Plasmas \textbf{24}, 092115 (2017).

\bibitem{Taniuti1969} T. Taniuti and N. Yajima, J. Math. Phys. \textbf{10}, 1369 (1969).

\bibitem{Chowdhury2017b} N. A. Chowdhury, A. Mannan, and A. A. Mamun,  Phys. Plasmas \textbf{24}, 113701 (2017).

\bibitem{Schamel2002} R. Fedele and H. Schamel, Eur. Phys. J. B \textbf{27}, 313 (2002).

\bibitem{Kourakis2005} I. Kourakis and P.K. Shukla, Nonlinear Proc. Geophys. \textbf{12}, 407 (2005).

\bibitem{Fedele2002} R. Fedele and H. Schamel, Eur. Phys. J. B \textbf{27}, 313 (2002).

\end{thebibliography}
\end{document}